# Electron-wave-stimulated mid-infrared emission from graphene-substrate quantum oscillators

Sunhwa Hong[1,3]‡, Moo Jin Kwak[2]‡, Yunseok Lee[3], Chan-Jin Kim[1,3], Sung Jin Hong[3], Ha Eun Lee[1,3], Yejun Lee[1], Koeun Kim[1], Juhyen Lee[2], Minkyung Lee[2], Youngdeog Koh[2], Joonhyun Lee[2], Miyoung Kim[3], Zee Hwan Kim[1], Myung Jin Park[3*], Hoon Wee[2*], Byung Hee Hong[1,3*] and Konstantin S. Novoselov[3,4]

[1]*Department of Chemistry, Seoul National University, Seoul 440-746, Korea.*

[2]*Samsung Research, Samsung Electronics, Seoul 06765, Republic of Korea*

[3]*Graphene Square Inc. & Graphene Research Center, Advanced Institute of Convergence Technology, Suwon 16229, Republic of Korea*

[4]*Department of Materials Science & Engineering, National University of Singapore, Singapore*

‡These authors contributed equally to this work.

**Generating tunable, high-intensity mid-infrared (MIR) to terahertz (THz) radiation on-chip remains a formidable challenge due to the rigid spectral limits of conventional thermal emitters[1]. While graphene has emerged as a promising platform for light-matter interaction, active control of its radiative properties has been largely confined to surface-limited phenomena mostly associated with plasmons[2-3]. Here, we introduce a new MIR radiation platform where multi-layer chemical vapor deposition (CVD) graphene is integrated with modular, vibrationally active dielectric substrates, ranging from organic thin films and inorganic matrices[4-5]. A pivotal discovery is that the long-range de Broglie wavelength of drift carriers enables coherent coupling with vibrational transition dipoles deep within the substrate bulk. This transforms the substrate into a three-dimensional volume emission source, where complex spectra of characteristic molecular and lattice vibration energies are additively combined on demand. The exponential scaling of radiation intensity ($\log I \propto V$) appears when the electrons' drift velocity in graphene exceeds the sound velocity of the substrates, consistent with quantum stimulated amplification**

**associated with Cerenkov electron-phonon instability[6,7]. Our work redefines the passive dielectric substrate as an active, programmable component driven by electron waves, paving the way for next-generation system-on-a-chip MIR-THz photonics[8], environmental and biomedical sensing, and highly efficient mode-specific electrothermal applications.**

The ability to actively manipulate MIR to THz radiation has advanced significantly through 2D nanophotonics[9-10], overcoming the classical thermodynamic constraints of conventional emitters. Recent breakthroughs have enabled the precise manipulation of spectral, polarization, and directional properties[11]. Notable examples include leveraging the phase transition of vanadium dioxide ($VO_2$) to dynamically switch from an omnidirectional low-emissivity state to a highly directional emission state[12], and utilizing flat band designs in distorted photonic lattices to eliminate the "rainbow effect," thereby achieving stable, ultra-narrowband thermal emission across wide viewing angles[13].

Despite these innovations, existing technologies face substantial hurdles such as complicated fabrication processes including electron-beam lithography[16]. Furthermore, the active control of graphene's radiative properties remains restricted to surface-confined modes, which suffer from small interaction volumes and low energy conversion efficiencies[2-5]. Ultimately, even with these nanophotonic enhancements, the fundamental emission intensity remains constrained by classical Joule heating and power-law scaling[1].

Here, we introduce a new effective MIR radiation platform that entirely bypasses the need for complex nanolithography or discrete phase-change materials by integrating multi-layer CVD graphene with modular dielectric molecular oscillators. This approach transitions from surface-limited transport to three-dimensional volume emission, driven by the non-equilibrium dynamics and wave nature of electrons. By stacking layers of different materials, individual spectral features can be combined into a single emission spectrum, while the bandwidth can also be tuned by adjusting the substrate thickness.

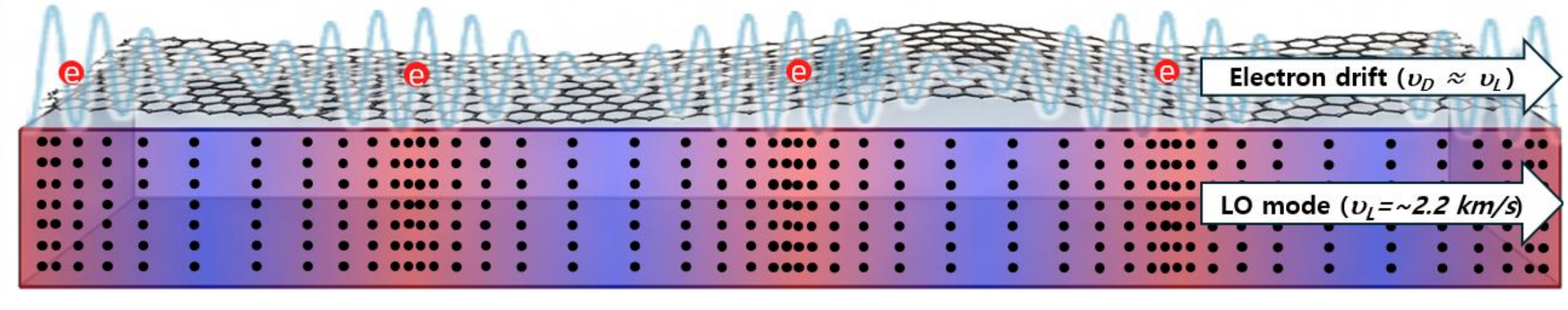

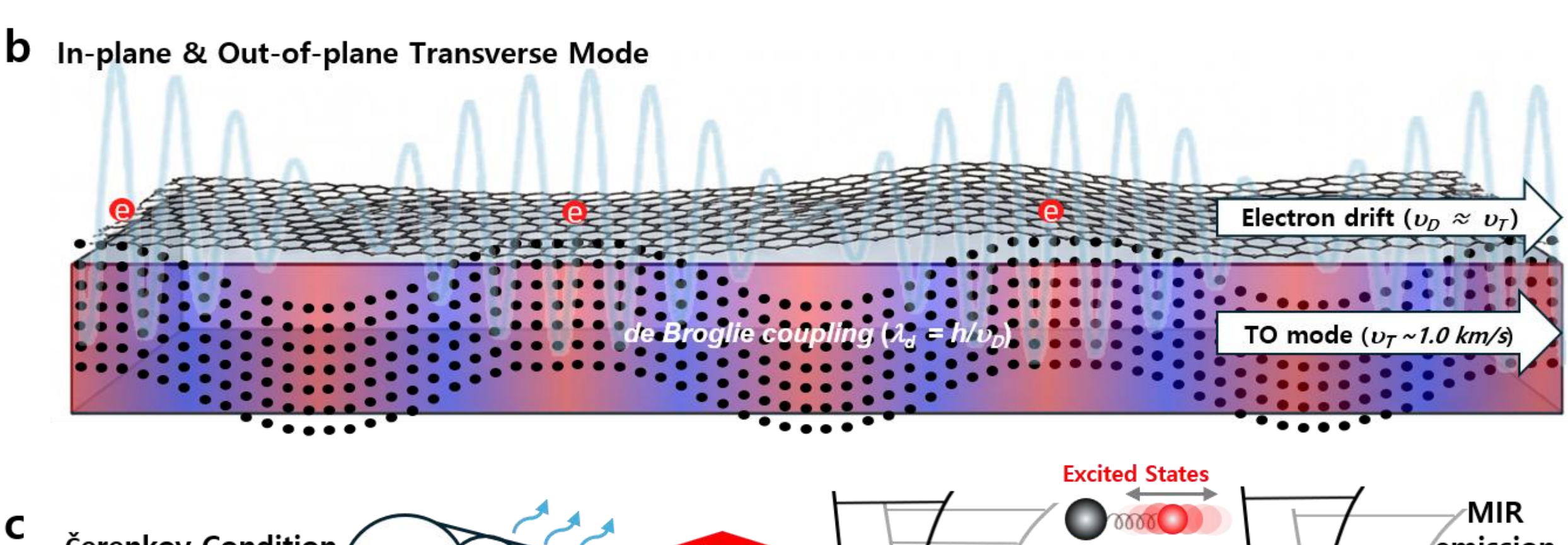

**Figure 1 | Origin of electron-phonon instabilities in graphene-substrate oscillators for MIR spectral emission. a** and **b**, Schematics of Čerenkov electron-phonon instability between 3-layer graphene and a polymer substrate (polyimide) with respect to longitudinal and transverse sound waves ($\upsilon_L$ and $\upsilon_T$), respectively. **c,** Schematic of Čerenkov condition that sonic boom like shockwaves coherently excite molecular or lattice vibrational modes of the substrates, followed by characteristic MIR emissions.

## Electron-phonon instabilities in graphene-substrate quantum oscillators

In our device, a modest bias of 17~40 V/cm results in carrier drift velocity ($\upsilon_D$) of approximately 850~4,000 m/s for charge carrier mobility of 5,000~10,000 $cm^2/Vs$ (Extended Data Fig.1), which is comparable to typical longitudinal and transverse sound velocity ($\upsilon_L$ and $\upsilon_T$) of conventional polymer substrates at room temperature. In polyimide (PI), the longitudinal wave travels at ~2.2 km/s, while the transverse wave moves at about half that speed (Fig. 1a, b). When the accelerated electron's drift velocity ($\upsilon_D$) matches the sound velocity of PI ($\upsilon_L$ or $\upsilon_T$), meeting the Čerenkov condition, coherent shockwaves excite wide range of high-lying vibrational modes in the substrate[14], followed by characteristic MIR emissions (Fig. 1c, Fig. 2a-b). In addition, the sound velocity of polymers decreases by up to ~40% at high temperatures close to glass transition ($T_g$)[15], which is expected to lower the onset bias for Čerenkov electron-phonon instability. It was also reported that the intrinsic supersonic nature of graphene's carriers ($\upsilon_F \approx 10^6$ m/s) allows the system to enter a regime of electron-phonon instability even for sub-sonic drift conditions at low bias[10]. The radiation intensities measured across the electrodes under AC and DC biases clearly show that more accelerated drift electrons result in stronger MIR emission (Extended Data Fig. 2), which also supports the proposed emission mechanism.

## Three-dimensional vibration dipole excitation *via* de Broglie coupling

A key conceptual breakthrough in this study is that electron-driven excitation extends beyond the interface into the substrate bulk, producing three-dimensional volumetric emissions *via* collective molecular oscillations. This is facilitated by the extended de Broglie wavelength ($\lambda_b$) of the drift carriers, which reaches the micrometer scale at low-bias conditions. For example, slowly drifting electrons (100~1,000 m/s) enables the penetration of their electric field deep into substrates (~0.7~7 μm). This long-range quantum wave allows the electronic wave packets to bypass immediate surface scattering and engage with vibrational dipoles deep within the substrate bulk, followed by the volumetric MIR emission corresponding to characteristic vibrational modes of a substrate on both sides (Fig. 2a-b, Extended Data Fig.3).

The hallmark of this process is the exponential scaling of the MIR emission intensity ($I$) with the bias voltage ($V$). Unlike classical Joule heating-dominated blackbody radiation, which follows a power-law relationship ($I \propto V^2$), our results exhibit a clear linear trend on a semi-logarithmic scale ($\log I \propto V$), confirming the presence of a quantum-driven stimulated process under Čerenkov conditions (Fig. 2c).

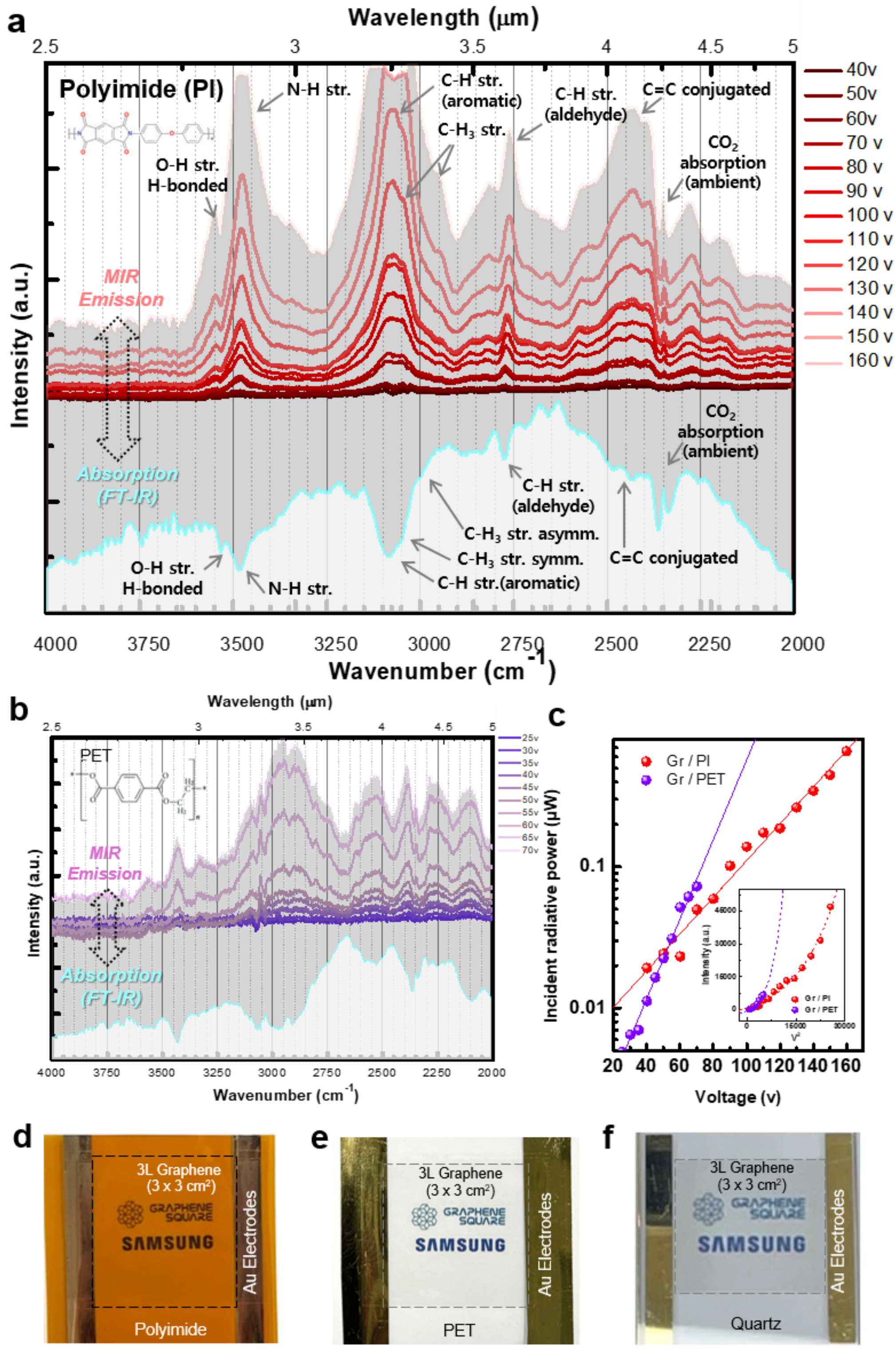

**Figure 2 | MIR emission characteristics of graphene-polymer Joule emitters. a** and **b,** Comparison between the Joule emission (upper) and FT-IR absorption (lower) spectra from graphene-PI and graphene-PET substrates, respectively, showing the identical peak positions corresponding to characteristic vibrational absorption. **c**, Radiation power (*P*) plots with respect to bias voltages (*V*), showing $\log P \propto V$ dependency. The inset shows the non-linear relationship between MIR intensity (*I*) and electric power ($\propto V^2$). **d-f**, Photographs of 3 x 3 cm$^2$ 3L graphene transferred on PI, PET and quartz substrates with thermally deposited Au electrodes, respectively.

## MIR emission spectrum on demand

This volumetric coupling allows for the additive spectral design of the output as shown in Fig. 3. By stacking modular layers of diverse polar dielectrics, such as PI, sapphire, quartz, and amorphous glass, the amplified phonons from the graphene 'engine' pump the specific molecular oscillators of each layer simultaneously. (The sapphire with no vibrational modes in MIR ranges plays a role as a blank substrate.) The resulting spectrum is a superposition of the vibrational signatures of the integrated materials, providing a platform for tailored spectral outputs that can be customized for any specific application. The engagement of the bulk volume is further evidenced by the downstream spatial intensification of the emission and its millisecond-scale temporal persistence post-bias, reflecting the collective relaxation of the bulk phonon manifold (Extended Data Fig. 4). While the electron drift in the graphene channel ceases instantaneously, the mechanical and vibrational energy already pumped deep into the low-thermal-conductivity polymer bulk takes significant time to dissipate *via* multi-phonon scattering[16].

Following the observation that the emission peaks coincide with substrate vibrational modes, we next examined whether the spectral output could be tailored through substrate and dielectric overlayer engineering. Figure 3a schematically shows the device structure, where a multilayer graphene Joule emitter on sapphire (*Blank*) and quartz (*Spectrum B*) substrates exhibits sparse and rich spectra, respectively, within the measured range. In addition, by stacking different dielectric overlayers such as PDMS (*Spectrum A*) and PI (*Spectrum C*) on top of the graphene Joule emitters, the overall spectra can be additively synthesized (*A only, A+B, A+C, A+B+C*, …) as shown in Fig. 3b–e. Thus, we can optimize the resonance of radiative energy transfer from the graphene Joule emitter to target objects (Extended Data Fig. 5 and 6). For example, designed MIR emission corresponding to -O-H, -N-H and -C-H vibrations are particularly important when heating carbohydrates, protein and fats, respectively (Extended Data

Fig. 7). This stands in clear contrast to the blackbody radiation of conventional coil heaters, which exhibit a broad spectrum spanning the near-IR to MIR range (Extended Data Fig.8).

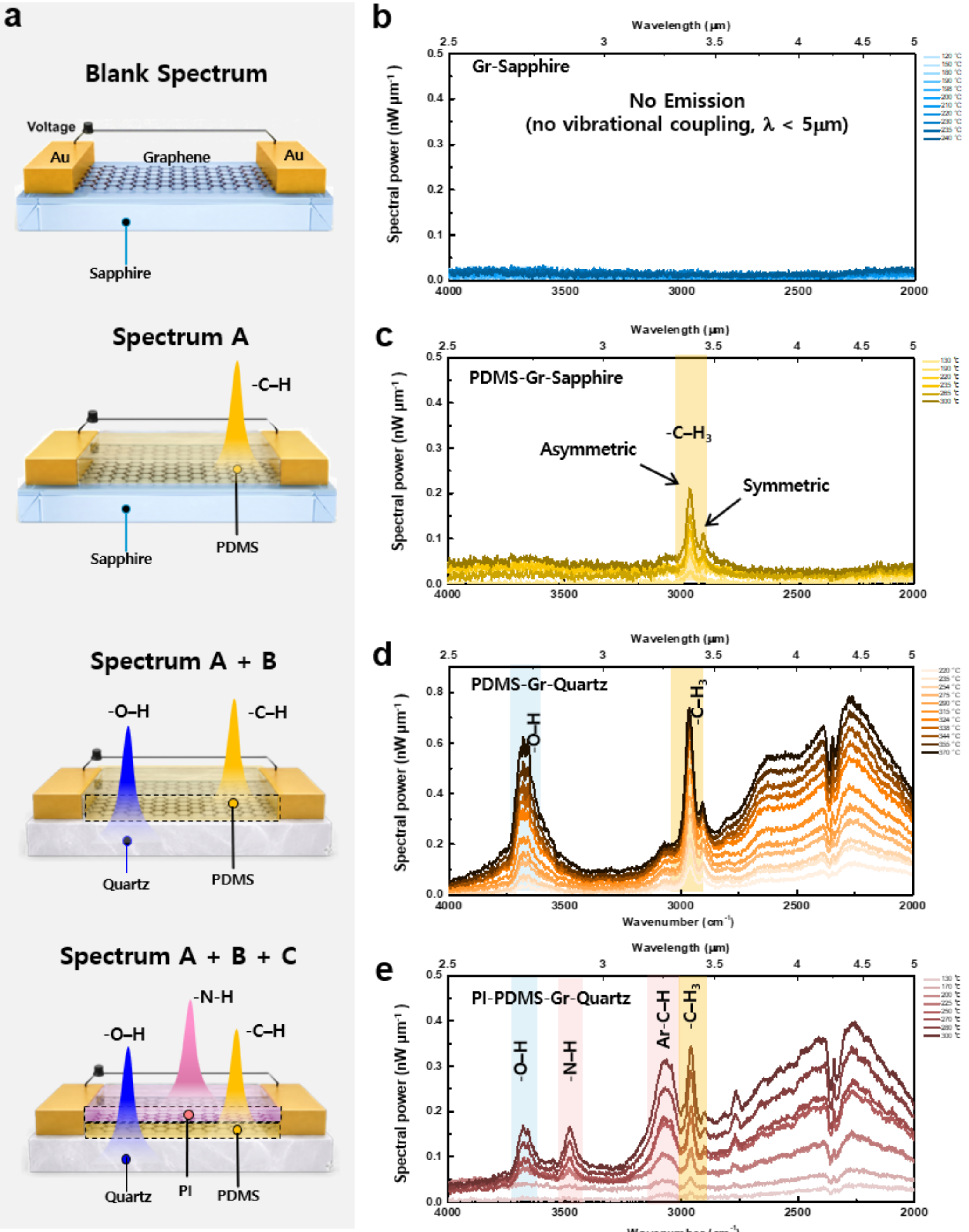

**Figure 3 | Materials-dependent characteristic Joule emission spectra and their additive combination. a**, Schematic illustration of an additive Joule emission device capable of incorporating distinct materials on its top and bottom surfaces. **b-d**, Joule emission spectra from graphene-quartz, PDMS-graphene-quartz, PI-graphene-quartz, and PI-PDMS-graphene-quartz substrates, respectively. The inset in b shows the MIR spectrum from PI-graphene-sapphire Joule heater, where the sapphire doesn't show any vibrational modes in 2.5~5 $\mu$m range.

## Thickness and crystallinity dependent spectral broadening

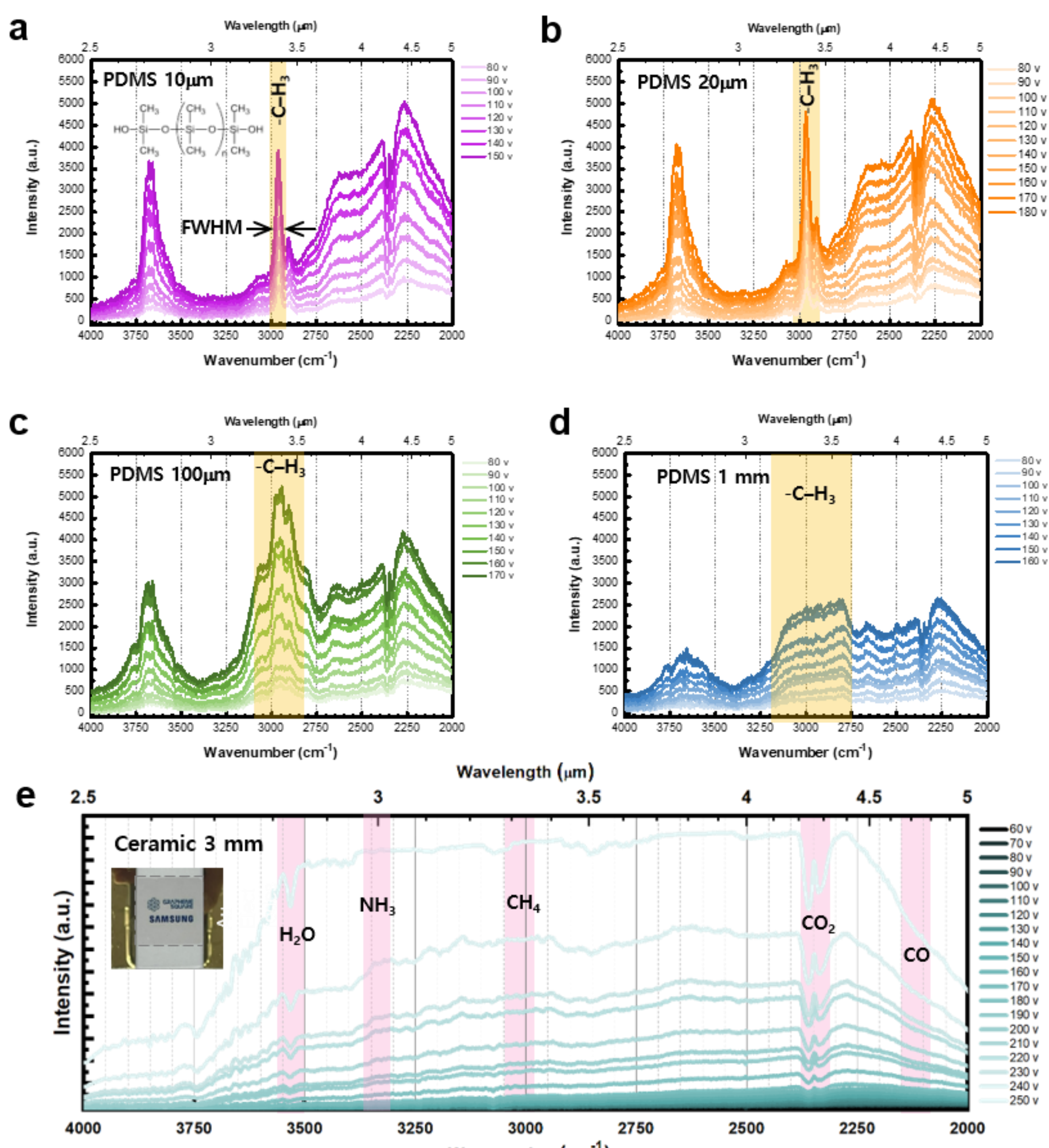

**Figure 4 | Thickness and crystallinity dependent broadening of MIR emission spectra from graphene-substrate quantum oscillators. a-d**, MIR spectra of 10 μm, 20 μm, 100 μm and 1mm thick PDMS placed on graphene-quartz Joule heaters, respectively. The FWHM doesn't change much for the thickness below 10 μm, indicating that the electron wave's penetration depth is as long as a few microns scale. **e**, Extraordinary broadband MIR spectrum is observed from graphene-ceramic (amorphous) emitters, which is useful as a source to detect various gas molecules. The absorption peak around 3,550cm$^{-1}$ and 2,370cm$^{-1}$ are from atmospheric $H_2O$ and $CO_2$. As a relatively low-temperature (~ 600 K) MIR source, its merit lies in covering the 2.5–5.0 μm (4000–2000 cm$^{-1}$) spectral range. The inset photograph shows 3 x 3 cm$^2$ 3L graphene transferred on ceramic glass substrate with thermally deposited Au electrodes.

Interestingly, the spectral profile is highly sensitive to the interaction volume. While thin substrates produce sharp vibrational peaks corresponding their IR absorption spectra (Fig. 4a and b), thick substrates (for example in polymers, exceeding 100 μm and reaching mm scales) exhibit a transition toward broad emission bands (Fig. 4c, d. This broadening signifies consistent with multimode vibrational participation within disordered polymer networks, potentially accompanied by reabsorption and re-emission processes that redistribute spectral energy across the bulk. Similarly, in thick ceramic glass, the higher sound velocity ($\upsilon_L \approx 5$ km/s) within the random molecular network-inherently due to its amorphous nature- leads to broadband emissions rather than sharp peaks. This indicates that the vibrational excitation is not limited to the interfacial area but propagates throughout the substrate volume. Consequently, as observed in the thickness-dependent spectra, the emission intensity continuously scales up from the 10 μm to the 100 μm under identical bias conditions. For extremely thick substrates (such as the 1 mm), likely due to localized self-absorption and scattering within the bulk polymer, the excited vibrational energy becomes broadly redistributed throughout the disordered polymer network, evolving into a multi-mode broadband spectral profile rather than exhibiting well-defined peaks.

## Applications of synthetic MIR spectral emission

The ability to construct complex, multi-band spectra additively opens new frontiers in mode-specific chemistry. Traditionally, the primary bottleneck in achieving successful mode-specific chemical reactions has fundamentally stemmed from rapid vibrational relaxation, which inevitably induces thermal randomization of the injected energy before the targeted reaction can take place, as well as a lack of sufficient systematic experimental demonstrations. Despite the notable advancements in MIR lasers, a highly suitable, accessible, and easily tunable MIR source dedicated to selective molecular vibrational excitation has remained elusive. By matching the radiation peaks to the precise absorption frequencies of targeted molecular bonds, we can selectively drive chemical reactions[17,18]. This precision makes our platform uniquely suited for advanced food processing (Extended Data Fig. 7)[19]. Furthermore, the programmable nature of this MIR-THz source, combined with its high intensity and low-field operation, positions it as a key technology for secure, high-bandwidth on-chip biomolecular diagnostics [20]. The extraordinary broadband MIR spectrum in Fig. 4e would be useful as a source for sensors quantifying the concentration of various gas molecules[21,22], where atmospheric $H_2O$ and $CO_2$ show clear absorption dips around 3,250 $cm^{-1}$ and 2,325 $cm^{-1}$, respectively.

The chemical and mechanical stabilities of graphene allow operating temperature as high as 500 °C and reliability for more than 1,000 cycles of 60 min heating. In addition, the graphene heater delivers more heat than conventional sheath or SUS heaters at the same distance (Extended Data Fig. 9). We also confirmed that the double-side graphene Joule emitter is more effective in generating MIR power than the single-side emitter (Extended Data Fig. 10). The temperature of the Joule emitter surface can be precisely monitored by using the red-shift of MIR peak positions, which is primarily attributed to thermal softening of C-H and N-H stretching modes (Extended Data Fig. 11).

Given the unique features of graphene heating platforms and recent advances in the roll-to-roll (R2R) industrial production of CVD graphene[23], our findings are expected to open a new era of energy-selective energy-efficient MIR heating technology that can significantly reduce power consumption in homes and industries, replacing traditional coil-heater technologies that have dominated for over a century.

## Conclusions

We have experimentally demonstrated that Joule-heated multilayer graphene integrated with modular dielectric oscillators serves as a platform for spectrally selective MIR emission, stimulated directly by the wave nature of electrons. The transition from surface-limited transport to three-dimensional volume emission is strictly governed by the Cerenkov electron-phonon instability, resulting in the exponential scaling of radiative intensity. Thus, unlike conventional blackbody emitters, these graphene-based heaters generate distinct material-dependent emission peaks that can be systematically tuned through dielectric composition and thickness. Beyond conventional resistive heating, these findings establish graphene-dielectric heterostructures as a versatile platform for electrically driven, wavelength-selective MIR emission, offering new opportunities for targeted thermal processing, energy-efficient heating, and infrared photonic technologies.

## Online content

Any methods, additional references, reporting summaries, source data, extended data, supplementary information, acknowledgements, peer review information; details of author contributions and competing interests; and statements of data and code availability are available at https://doi.org/...

## Methods

### Fabrication of Joule emitter devices

A copper (Cu) foil (10 × 10 $cm^2$) was loaded into a quartz tube furnace and heated to 1,000 ºC under a flow of 5 sccm hydrogen ($H_2$) for 100 min. Graphene growth was then carried out using 50 sccm methane ($CH_4$), 5 sccm $H_2$, and 100 sccm argon (Ar) for 30 min. After growth, the graphene film on the Cu foil was laminated onto a thermal-release tape by applying soft pressure (0.2 MPa) between two rollers. The Cu foil was subsequently removed using a copper etchant, and the graphene film supported on the tape was rinsed with deionized water to remove residual etchant. The graphene/tape assembly was then brought into contact with a target substrate and passed through rollers while applying mild heat (90-120 ºC), resulting in transfer of the graphene film onto the substrate. By repeating this transfer process on the same substrate, multilayer graphene films were prepared[8]. Ti/Au electrodes (10nm/150 nm) were deposited by thermal evaporation. A lateral bias applied across the graphene channel generated current flow and Joule heating during device operation[6]. Optical photographs of the fabricated graphene heaters are provided in Fig. 2d-f.

**MIR emission measurements**

Voltage was applied to the fabricated electrodes using a power supply, and the resulting emission spectrum was measured in the mid-infrared (MIR) range using an NLIR S2050 Mid-Infrared Spectrometer.

The measured spectral signal was converted to wavelength-dependent radiative power, $P_\lambda(\lambda)$(W μm$^{-1}$), using the instrument calibration factor. All radiative quantities reported in this study were calculated from the calibrated power spectra within the wavelength range. The incident radiative power ($P_{inc}$) from the heater was calculated as:

$$P_{\mathrm{inc}} = \int_{\lambda_1}^{\lambda_2} P_\lambda\,(\lambda)\,d\lambda$$

where $P_\lambda(\lambda)$is the spectral radiative power of the emitter, and $\lambda_1$and $\lambda_2$correspond to wavelength ranges, respectively. To estimate the absorbed radiative power ($P_{abs}$) of a receiver, the emitter spectrum was weighted by the wavelength-dependent absorptance of the receiver, $\alpha_{\mathrm{rec}}(\lambda)$, and integrated over the same spectral range:

$$P_{\mathrm{abs}} = \int_{\lambda_1}^{\lambda_2} P_\lambda\,(\lambda)\alpha_{\mathrm{rec}}(\lambda)\,d\lambda$$

The wavelength-dependent absorptance of the receiver, $\alpha_{\mathrm{rec}}(\lambda)$, was estimated from transmission measurements as:

$$T(\lambda) = \frac{I(\lambda)}{I_0(\lambda)}, \alpha_{\mathrm{rec}}(\lambda) = 1 - T(\lambda)$$

where $I(\lambda)$ and $I_0(\lambda)$ are the transmitted intensities measured with and without the receiver, respectively. Reflection losses were neglected in this analysis. Accordingly, $\alpha_{\mathrm{rec}}(\lambda)$was used as a comparative absorptance metric for evaluating relative radiative coupling within the MIR range. The effective radiative absorption efficiency ($\eta_{\text{abs}}$) was then defined as:

$$\eta_{\text{abs}} = \frac{P_{\text{abs}}}{P_{\text{inc}}}$$

where $\eta_{\text{abs}}$represents the fraction of incident radiative power that can be absorbed by the receiver within the selected MIR range. This metric reflects spectral overlap between the emitter and receiver and does not include geometric view factors or non-radiative heat-transfer pathways such as conduction and convection. The accumulated thermal energy of the receiver, $Q$, was estimated from the temperature rise using:

$$Q = mc\Delta T$$

where $m$ is the mass of the receiver, $c$ is the specific heat capacity, and $\Delta T$ is the temperature rise relative to the initial state. The corresponding heat capacities ($mc$) of the ceramic and quartz receivers were 14.9 and 11.1 cal $^{o}C^{-1}$, respectively.

**THz carrier mobility measurements**

The carrier mobility of the 3L graphene on a substrate was measured by a non-destructive Terahertz Time-Domain Spectroscopy (THz-TDS) system (ONYX, das-Nano). The measurement was performed over a total mapping area of *25* x *25 mm*$^{2}$, with a spatial mapping resolution (step size) of 0.5 mm. The data was acquired using the system's minimum acquisition time of 25 ms per point.

## Data availability

Data are publicly available on Zenodo at https://doi.org/.....

zenodo……..

---

**Acknowledgements** The research leading to these results has received partial funding from Samsung Research under the Project for the Development of Graphene Heat Source Technology (Project No. RAJ0125ZZ-71RF).

**Author contributions** B.H.H proposed original idea and contributed to design of experiment, supervision, data analysis, writing, editing, funding acquisition, and led the discussions with participation from all co-authors. S.H. and M.J.K. contributed to data acquisition, data analysis, writing, editing, and the development of theoretical frameworks under the guidance of B.H.H., M.J.P. and H.W. Y.L. contributed to data acquisition. C.K. contributed to editing. M.K.L contributed to editing. Z.H.K contributed to data analysis. M.J.P contributed to design of experiment, data analysis, editing, writing, and funding acquisition. H.W contributed to supervision, and funding acquisition. S.H. and M.J.K. contributed equally.

**Competing interests** The authors declare no competing interests.

## Additional information

**Supplementary information** The online version contains supplementary material available at https://doi.org/.....

**Correspondence and requests for materials** should be addressed to Byung Hee Hong, Hoon Wee or Myung Jin Park or Hoon Wi (E-mails: byunghee@snu.ac.kr, hoon.wee@samsung.com, hanson2525@graphenesq.com)

**Peer review information** ... thanks …. anonymous, reviewer(s) for their contribution to the peer review of this work. Peer reviewer reports are available.

**Reprints and permissions information** is available at http://.....

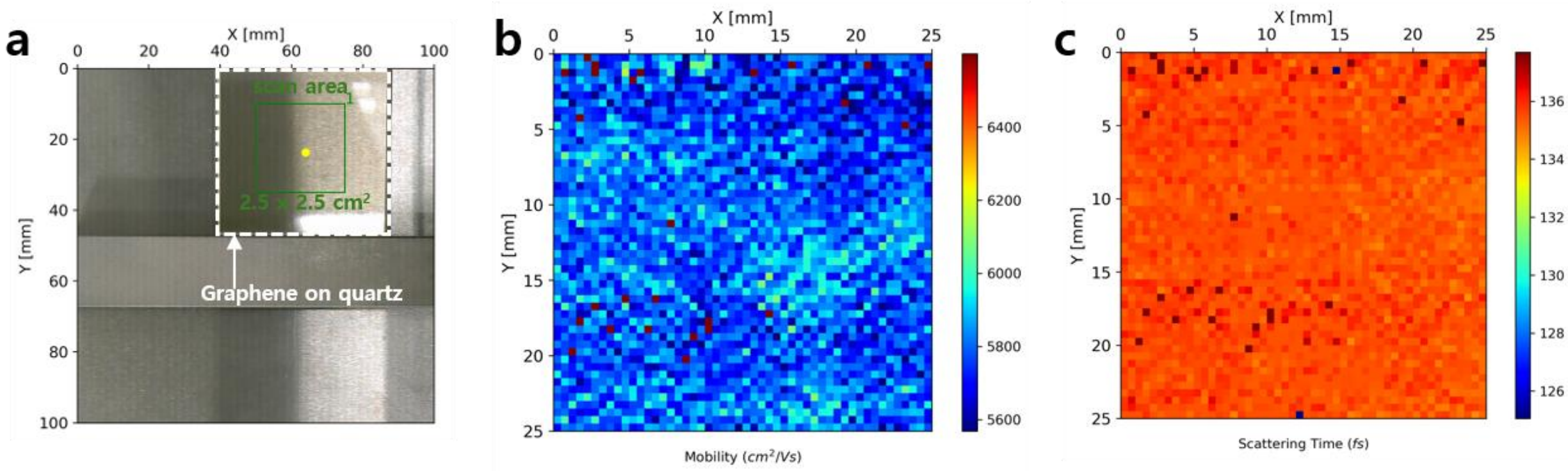

**Extended Data Fig. 1 | THz time-domain spectroscopy (THz-TDS) mapping of the carrier mobilities and scattering time of multi-layer CVD graphene on a ceramic glass substrate.** **a**, A photograph of the graphene-ceramic sample placed on a THz-TDS scanning stage. **b** and **c**, THz-TDS mapping of carrier mobilities and scattering time (fs scale) of 3L graphene on a ceramic substrate. The scattering time of ~135 fs indicates the high quality of 3-layered CVD graphene.

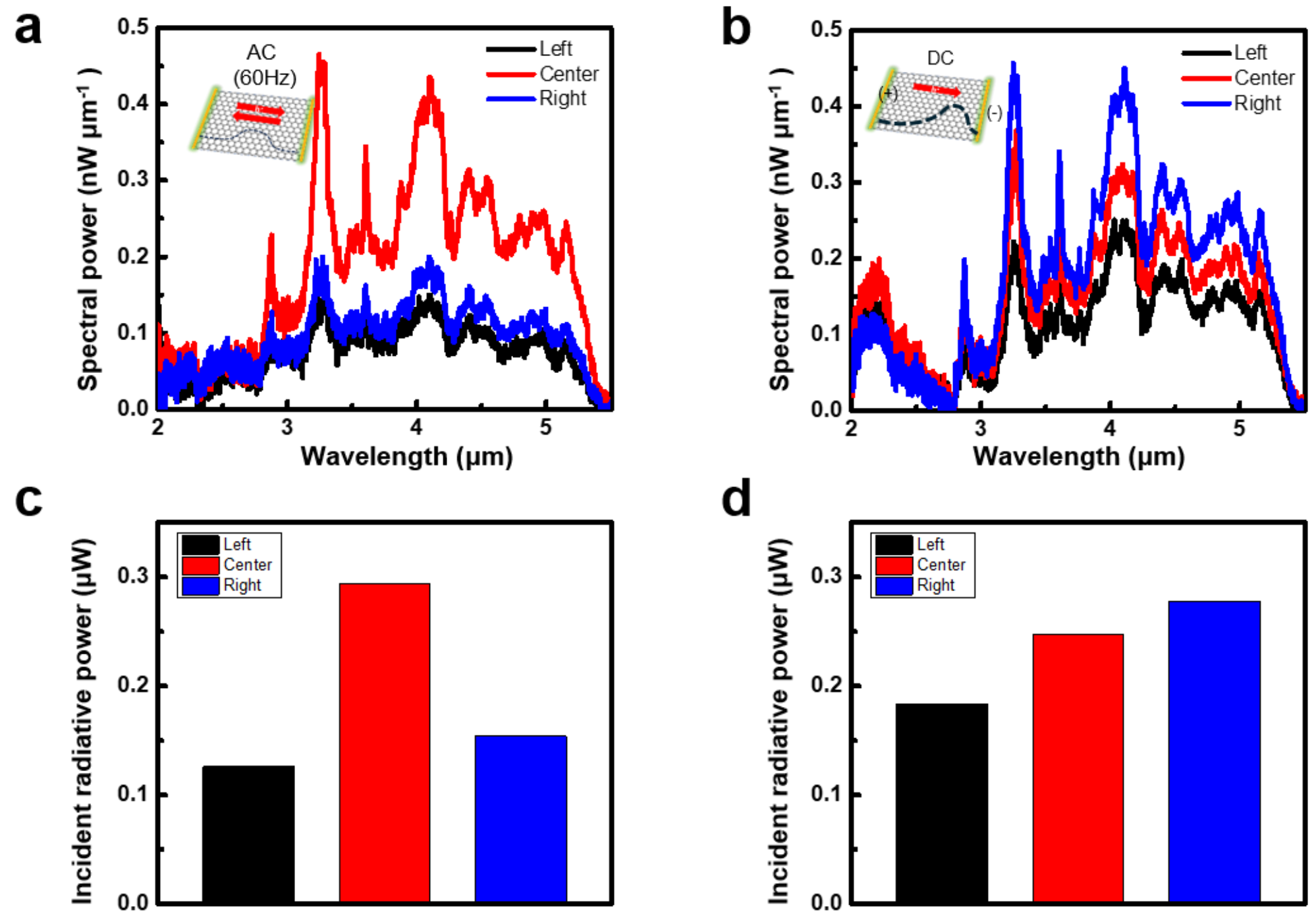

**Extended Data Fig. 2 | Mid-IR emission spectra and radiation power comparison between DC and AC powered graphene/PI heaters with respect to the position across the electrodes**, **a** and **b**, As drifting charge carriers are accelerated by electric field, the emission increases from left to right for DC. However, in AC bias, the maximum intensity of the mid-IR radiation is obtained at the centre region. **c** and **d**, Radiation power also increases from left to right for DC while, in AC bias, the maximum intensity of the mid-IR radiation is obtained at the center region.

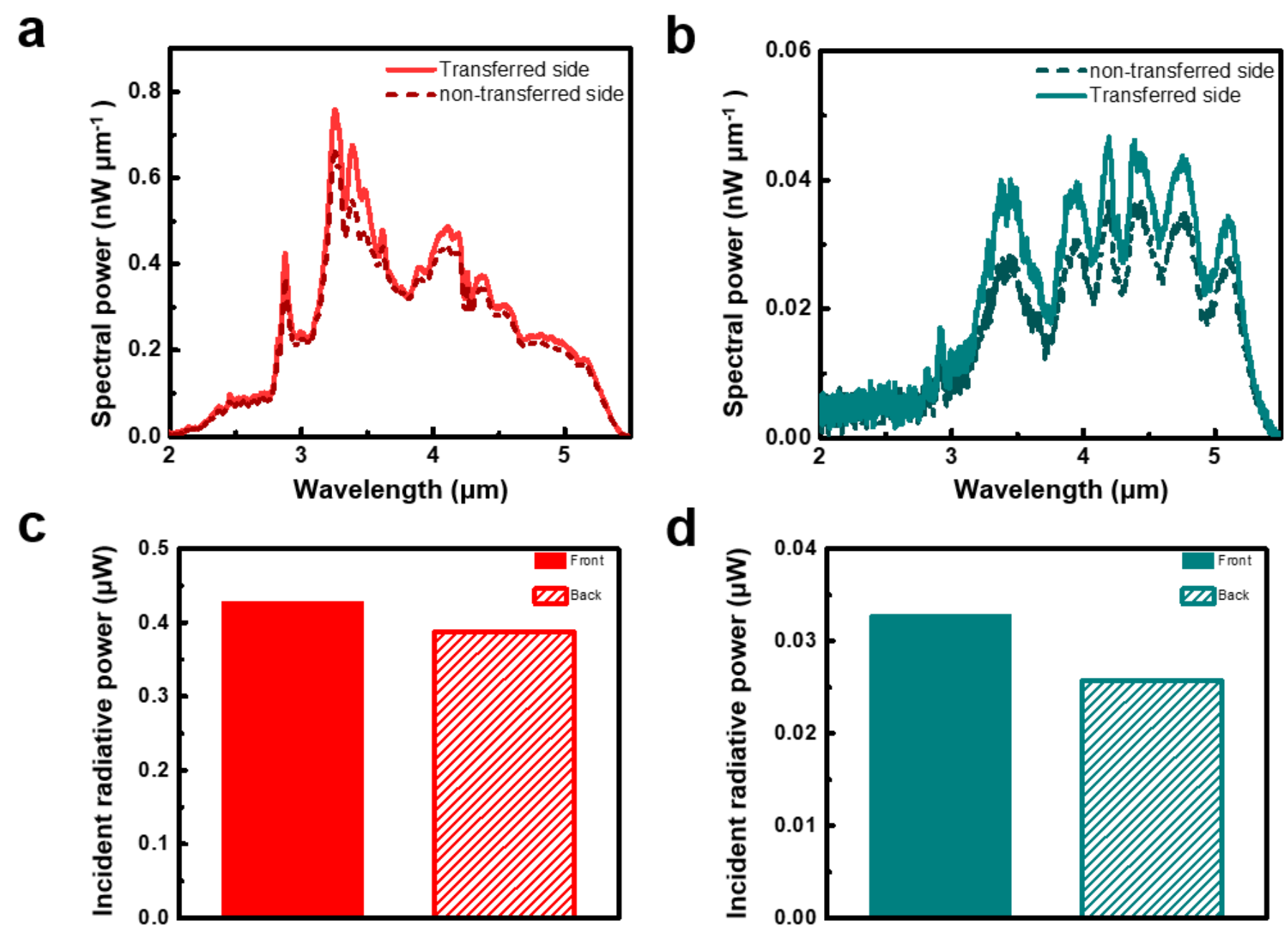

**Extended Data Fig. 3 | Emission spectra and radiation power comparison between the graphene-transferred side and the back side for substrates. a** and **b**, MIR spectra from both sides of PI and PET, respectively. **c** and **d,** Radiation power comparison between 2~5 mm MIR spectra from both sides of PI and PET, respectively.

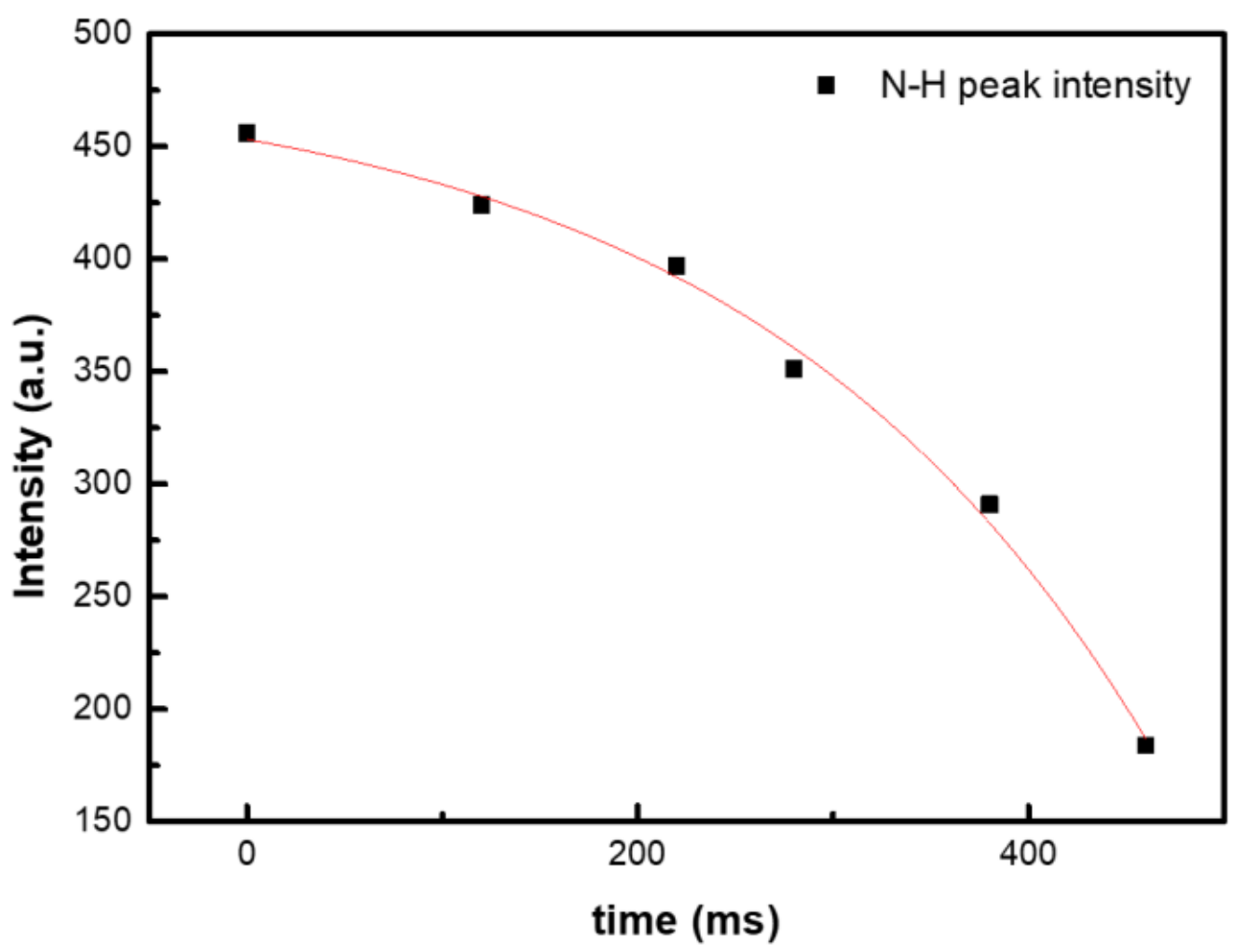

**Extended Data Fig. 4 | Decay of MIR peak intensities measured from a graphene-PI Joule oscillator after turning off bias voltage**. Upon removing the bias voltage, the specific MIR emission peaks exhibit a macroscopic relaxation time of ~500 ms before completely decaying. This prolonged persistence is orders of magnitude slower than typical electronic or surface-plasmonic relaxation times. We attribute this temporal delay to the collective relaxation of the bulk phonon manifold.

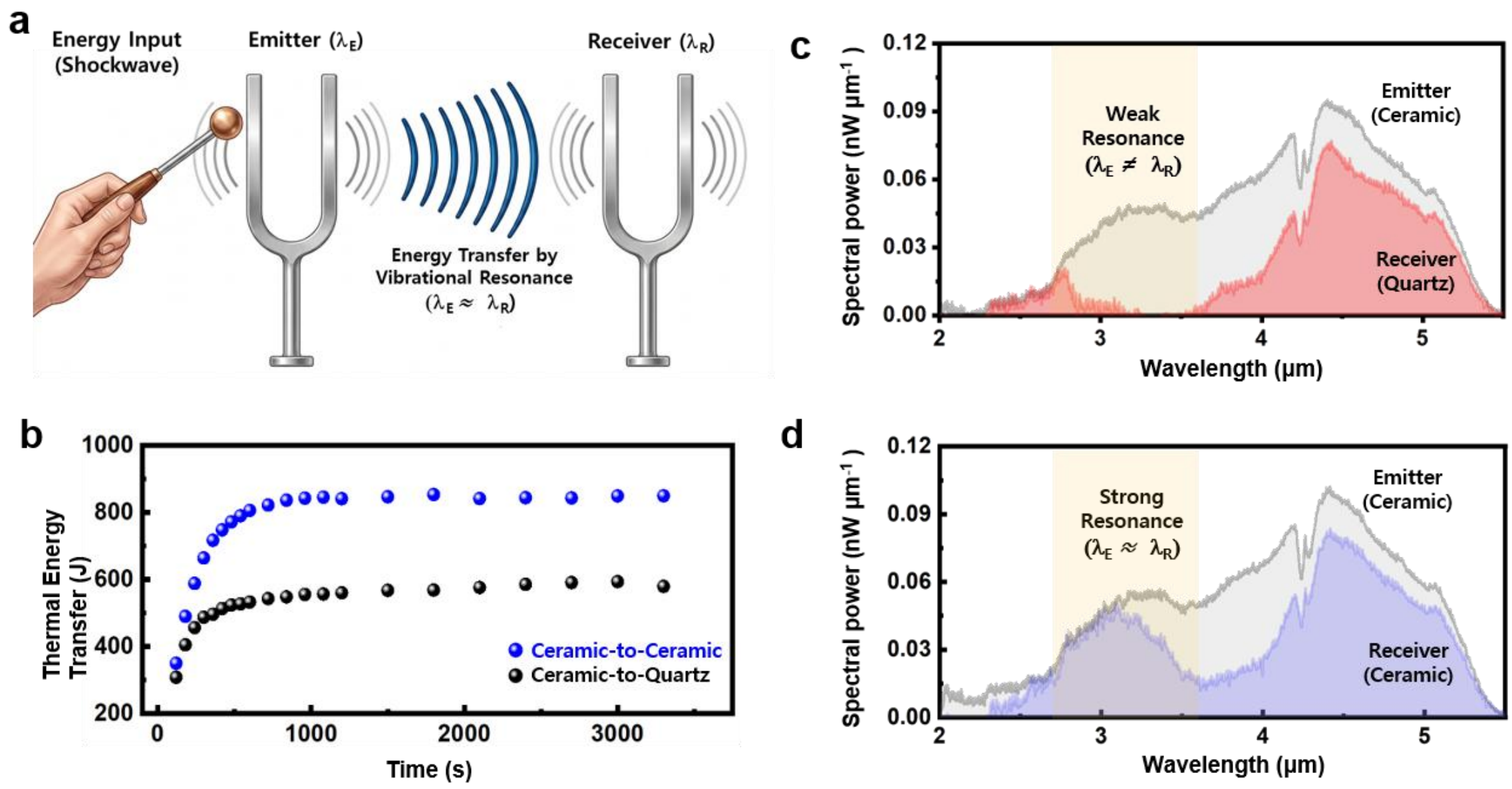

**Extended Data Fig. 5 | Resonant heat energy transfer between MIR emitters and receivers. a**, Schematic of resonance energy transfer between an emitter and a receiver. **b**, Thermal energy transferred from emitters (ceramic) to receivers (ceramic or quartz), where ceramic-to-ceramic resonance looks more efficient than ceramic-to-quarts. **c, d**, Spectral overlap of emitters and receivers in weak resonance and strong resonance, respectively.

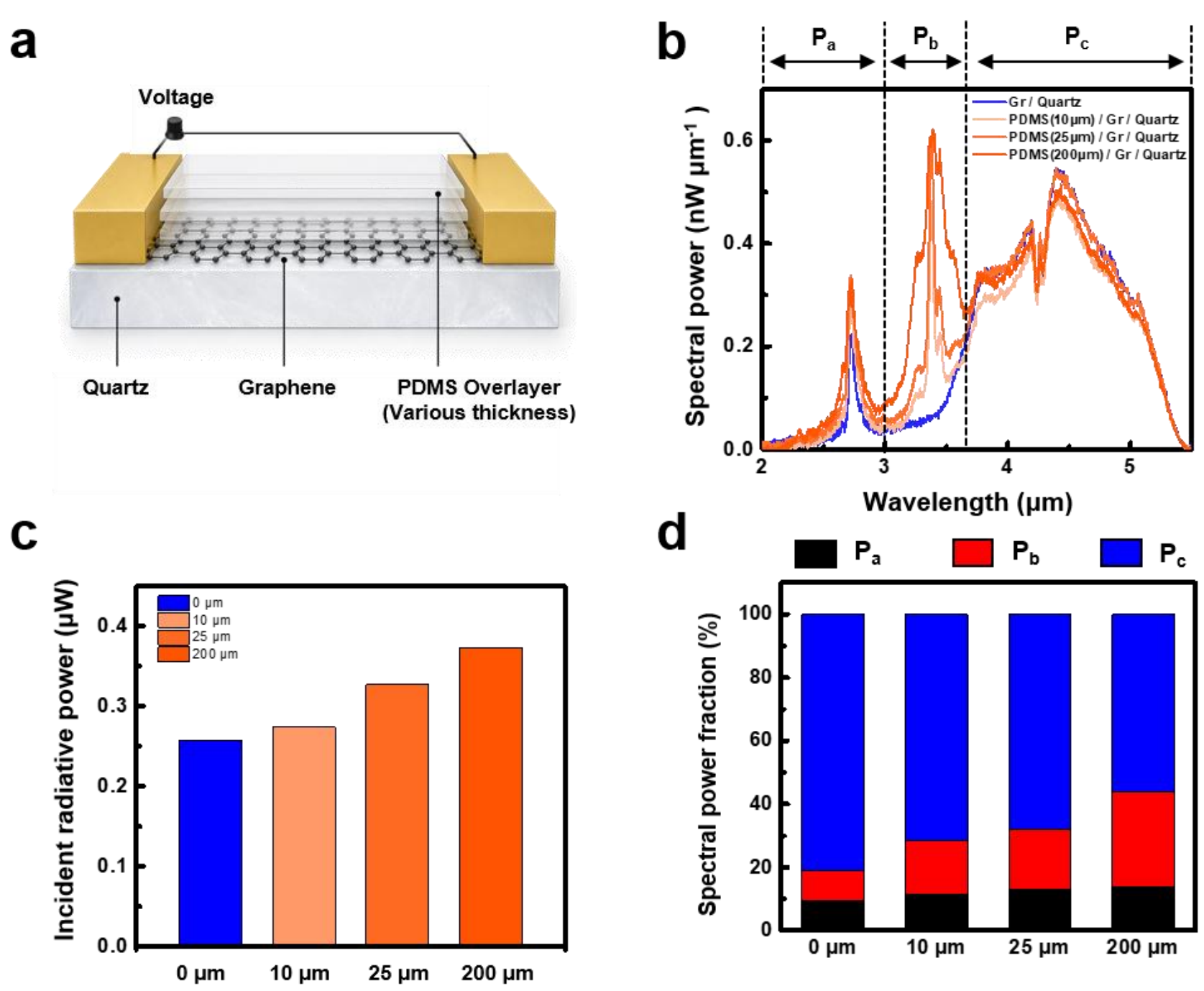

**Extended Data Fig. 6 | Thickness-dependent MIR radiation power of graphene-substrate Joule emitter. a**, Schematic of an overlayer-graphene-substrate Joule heater. **b**, MIR spectra broadening with increasing overlayer thickness. **c, d**, Radiation power and spectral power fraction with respect to overlayer thickness, respectively. The thicker overlayer provides broader emission and better radiative energy transfer.

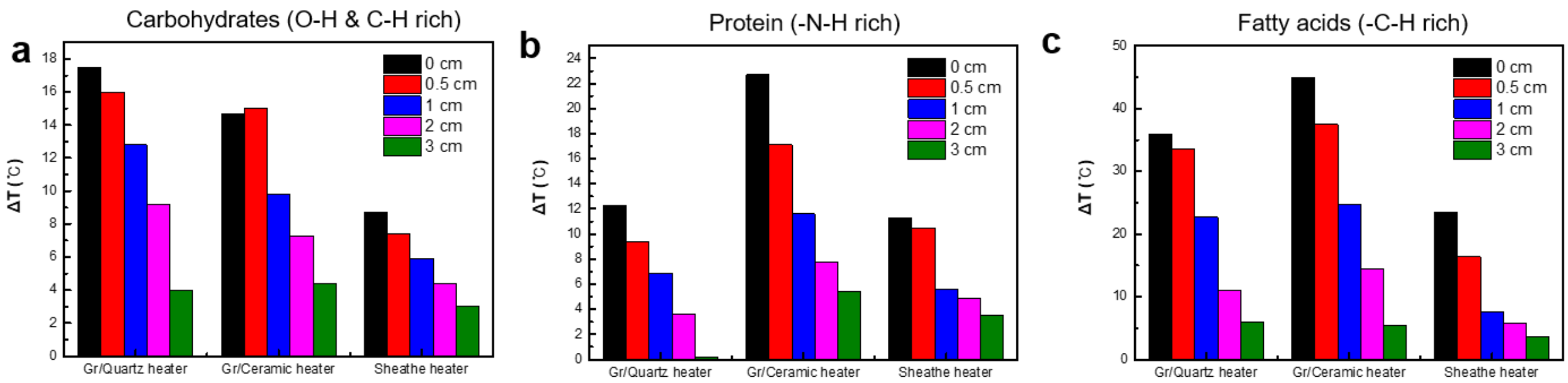

**Extended Data Fig. 7 | Temperature raised by different heaters and their depth profile for various food ingredients. a,** Temperature rise depth profile for –O-H rich carbohydrates (potato) heated by graphene-quartz, graphene-ceramic and sheathe heaters. **b**, Temperature rise depth profile for –N-H rich protein (chicken breast) heated by graphene-quartz, graphene-ceramic and sheathe heaters. **c**, Temperature rise depth profile for –C-H rich fatty acids (pork fat) heated by graphene-quartz, graphene-ceramic and sheathe heaters.

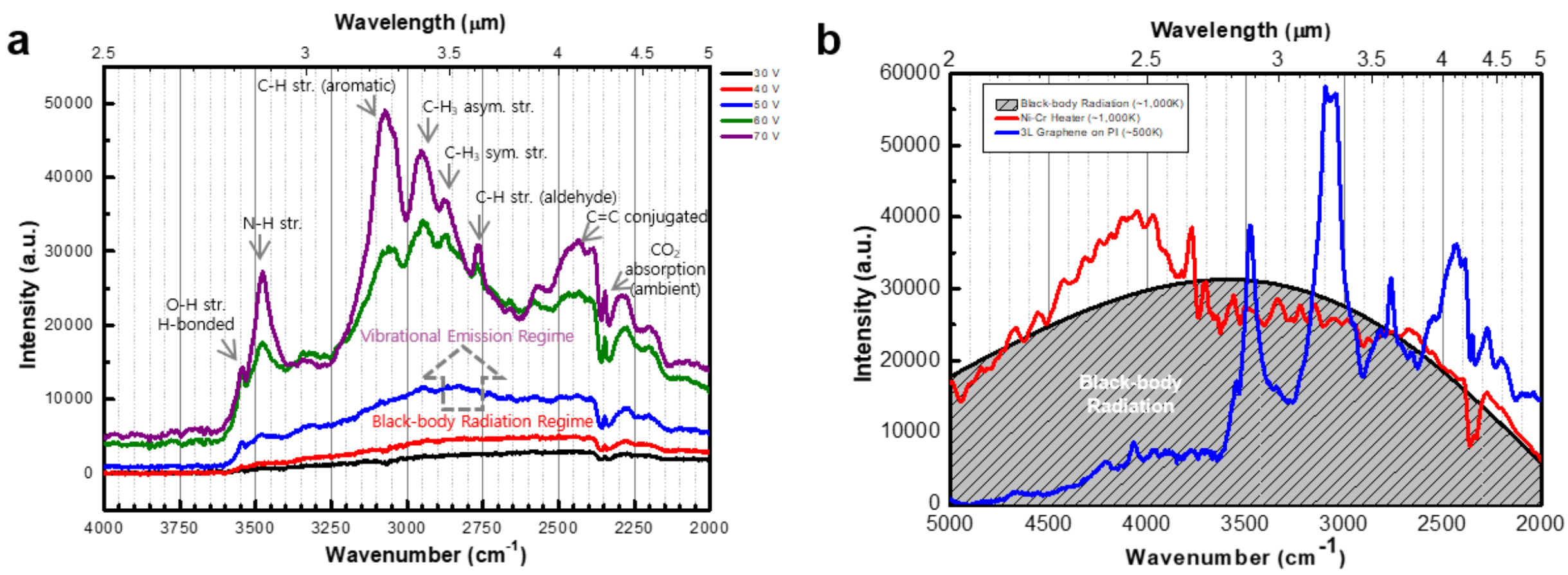

**Extended Data Fig. 8 | Mid-IR spectrum assignment and comparison with a blackbody heater (Ni-Cr). a**, Spectral as5ignment of the mid-IR radiation from graphene/PI, showing strong C-H stretching and C=C conjugation peaks. **b**, Comparison between graphene/PI and Ni-Cr heaters with similar power consumption (~300W). The Ni-Cr heater shows typical blackbody-like radiation, while the graphene/PI heater exhibits discrete molecular vibrational emission. The Ni-Cr heater (black, 1000 K) follows Planck's law, whereas the graphene heater (red, 500 K) shows sharp vibrational peaks deviating from the 500 K blackbody curve.

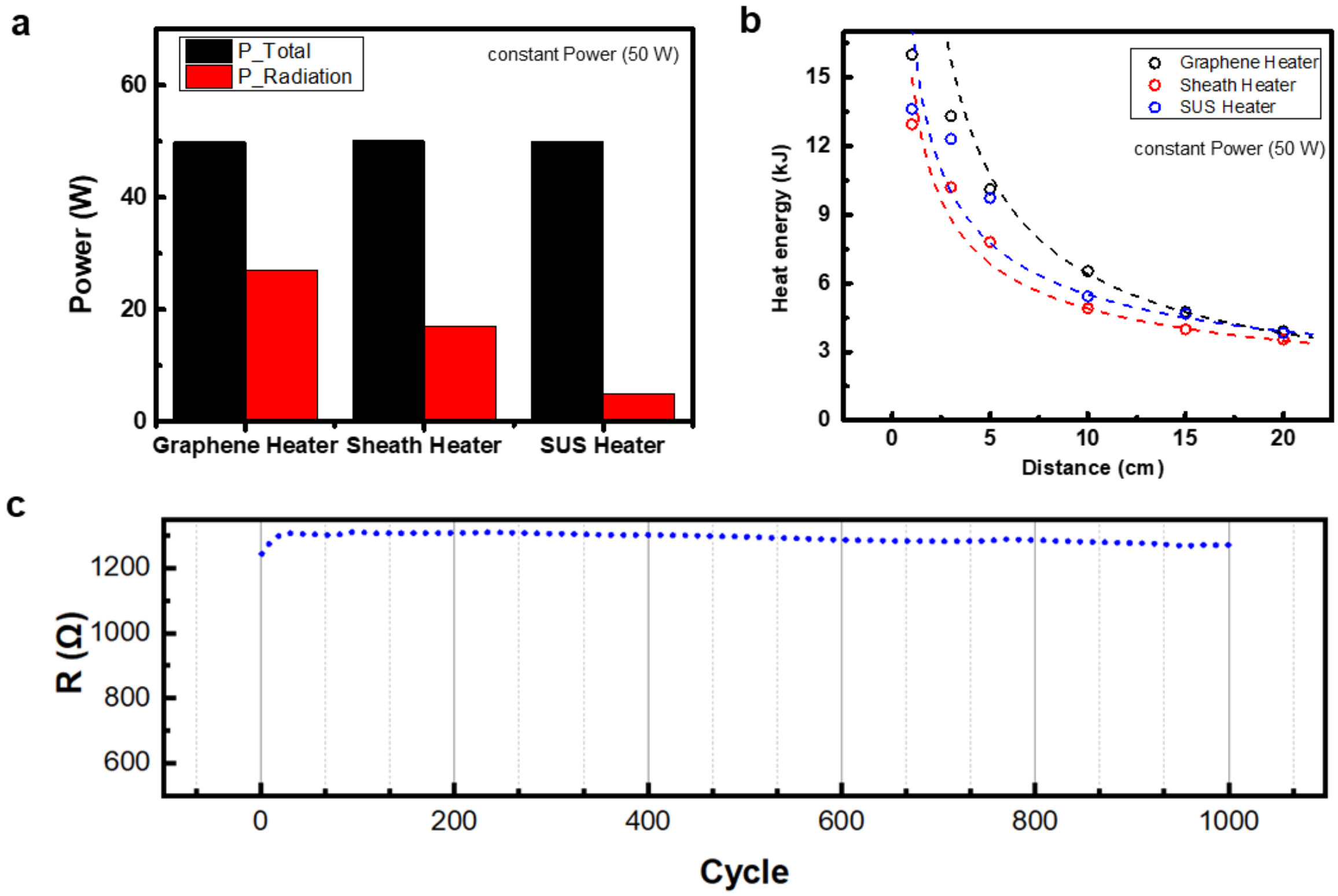

**Extended Data Fig. 9 | Radiative contribution of thermal energy generated by the graphene heater compared with conventional heaters. a,** Radiative power contribution of different heat sources compared at the constant total power (50 W)The heater dimensions are identical (135 x 150 mm$^2$). **b,** The radiative thermal energy absorbed by a blackbody-like black-coated aluminum plate with respect to the distance between the source and detector in a vacuum chamber (excluding conductive and convective heat transfer) at the constant power (50 W) The dashed lines are fitted by inverse-square-law. **c,** Resistance change with respect to the 1,000 cycles of 60 min heating and 20 min cooling, showing outstanding long-term reliability. The resistance remained constant within ±3% throughout the test, demonstrating excellent durability and operational stability under prolonged cycling.

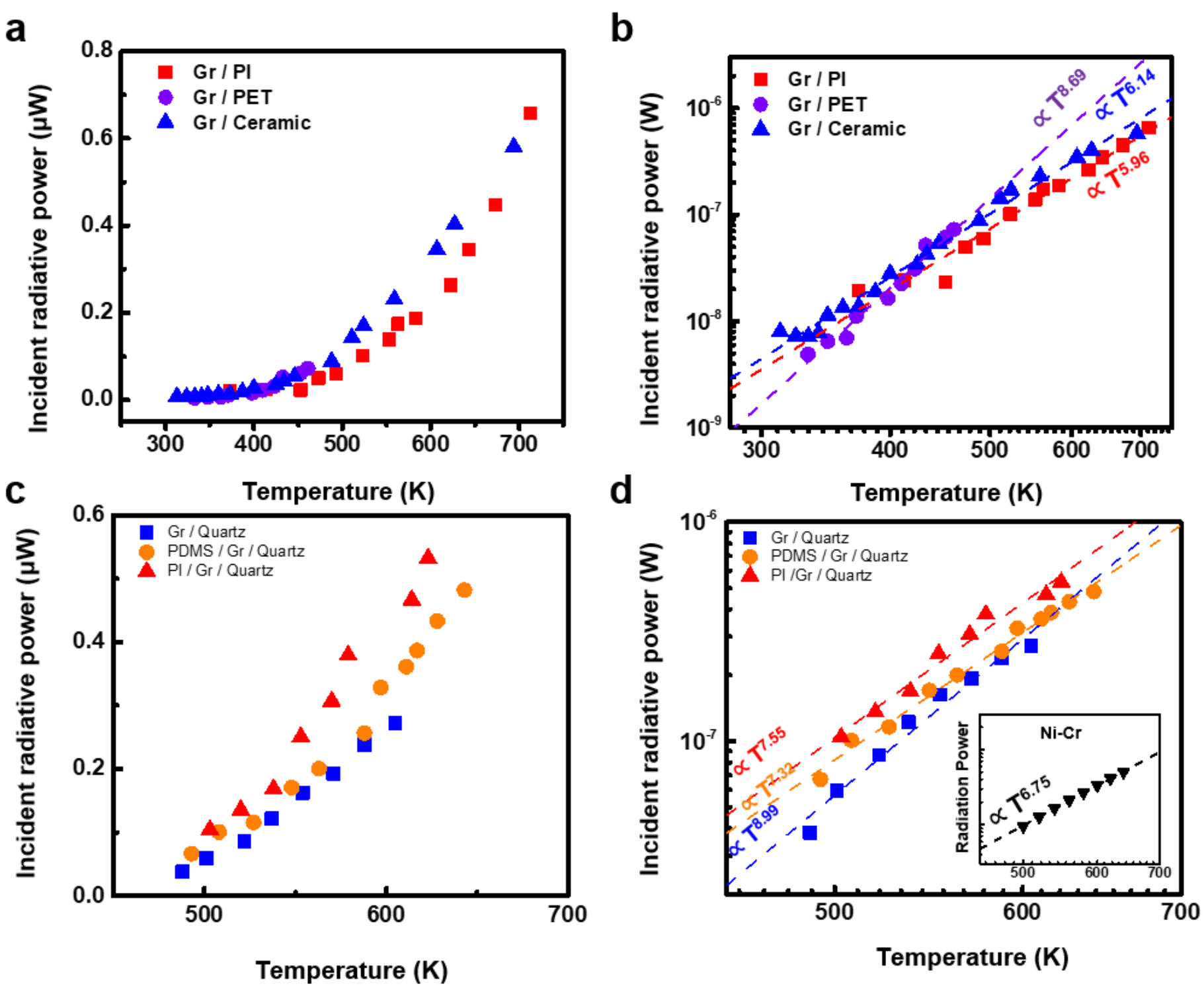

**Extended Data Fig. 10 | Radiative power with increasing temperature for single-side and double-side graphene-substrate interfaces. a** and **b**, Temperature dependence of radiation power for single-side graphene-substrate interface, proportional to $T^{5.96}$, $T^{6.14}$ and $T^{8.69}$ for graphene-PI, graphene-ceramic and graphene-PET, respectively. **c** and **d**, Temperature dependence of radiation power for double-side graphene-substrate interface, proportional to $T^{7.32}$, $T^{7.55}$ and $T^{8.99}$ for PDMS-graphene-quartz, PI-graphene-quartz and graphene-quartz, respectively. The inset in d shows the radiation power of Ni-Cr heaters (blackbody radiation), proportional to $T^{6.75}$.

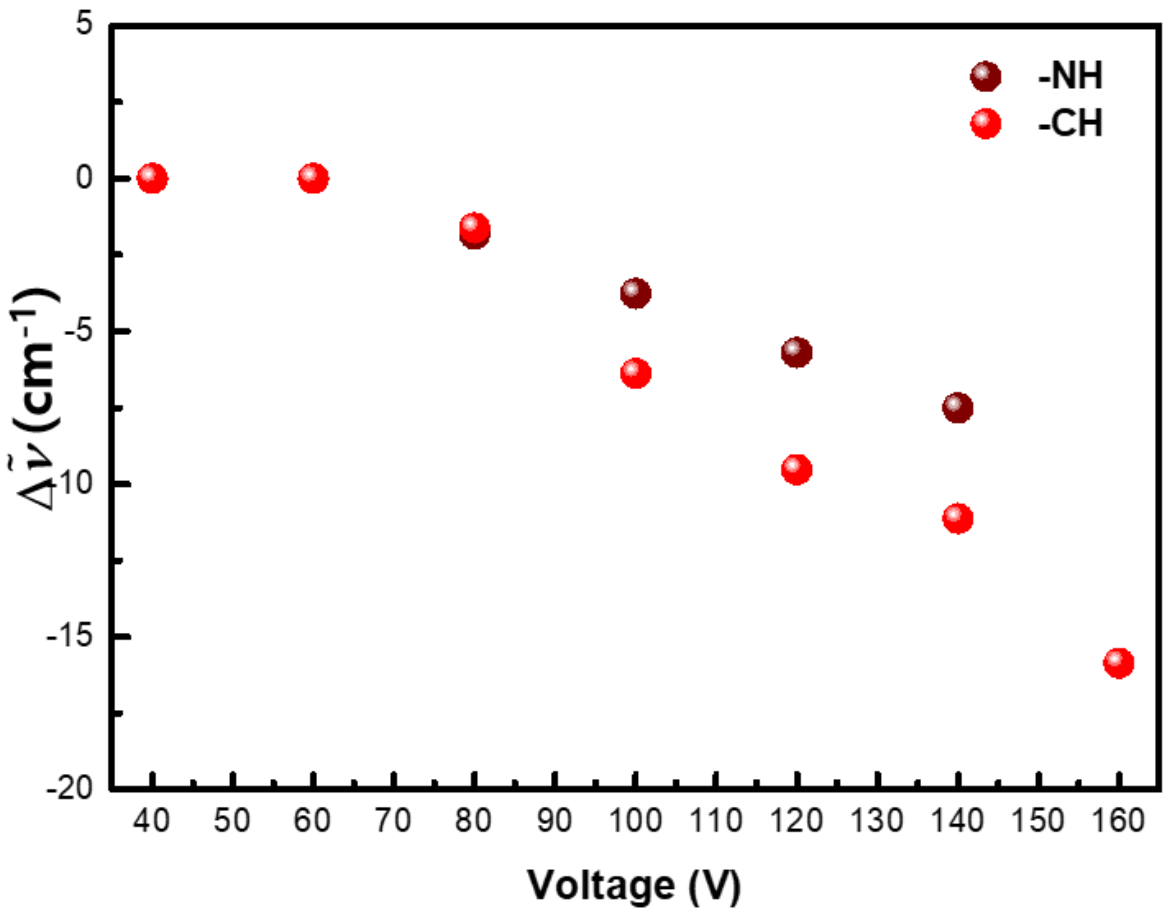

**Extended Data Fig. 11 | Red-shift of MIR peak positions with respect to bias voltage in graphene-PI oscillators.** As bias and temperature increase, the C-H vibrational emission peak exhibits a distinct red-sift from 2776 $cm^{-1}$ to 2759$cm^{-1}$, which is primarily attributed to thermal softening of C-H stretching modes. N-H vibration shows smaller red-shift, i.e., the lower temperature-frequency coefficient.